\documentclass[journal,twoside]{IEEEtran}
\pdfoutput=1
\usepackage{graphicx}
\usepackage{newtxtext}
\usepackage{newtxmath}
\usepackage{amsmath}
\usepackage{bm}
\usepackage{cite}
\usepackage[hidelinks]{hyperref}
\usepackage{eso-pic}
\usepackage{xcolor}
\usepackage{booktabs}
\usepackage[figuresright]{rotating}
\usepackage{enotez}
\setenotez{list-name=Errata,counter-format=alph}

\newtheorem{proof}{Proof}

\graphicspath{{./gfx/}}

\newcommand{\D}{\displaystyle}
\newcommand{\Vector}[1]{\bm{#1}}  
\newcommand{\Matrix}[1]{\bm{#1}}  
\newcommand{\Transpose}{\mathrm{T}}  
\newcommand{\refEq}[1]{(\ref{#1})}               
\newcommand{\refFig}[1]{Fig.~\ref{#1}}           
\newcommand{\refFigBegin}[1]{Figure~\ref{#1}}    
\newcommand{\refSec}[1]{Sect.~\ref{#1}}          
\newcommand{\refTable}[1]{table~\ref{#1}}        

\hypersetup{%
    pdfauthor={Gernot Herbst},
    pdftitle={Transfer Function Analysis and Implementation of Active Disturbance Rejection Control},
    pdfcreator={},pdfproducer={}
}

\begin{document}

\AddToShipoutPicture*{%
  \AtPageUpperLeft{%
    \setlength\unitlength{1cm}%
    \put(0,-0.5){\begin{minipage}[c]{\paperwidth}
    \footnotesize\centering\textcolor{black!50}{%
    This is a preprint of an article published in \emph{Control Theory and Technology}.
    The final authenticated version is available online at:} \ \textcolor{blue!60}{\url{https://doi.org/10.1007/s11768-021-00031-5}}%
    \end{minipage}}%
  }
  \AtPageLowerLeft{%
    \setlength\unitlength{1cm}%
    \put(0,+0.5){\begin{minipage}[c]{\paperwidth}
    \footnotesize\centering\textcolor{black!50}{%
        This revision of the preprint includes a post-publication correction.%
    }%
    \end{minipage}}%
  }
}

\title{Transfer Function Analysis and Implementation of Active Disturbance Rejection Control}

\author{Gernot Herbst\ \href{https://orcid.org/0000-0002-4638-5378}%
    {\raisebox{-0.3pt}{\includegraphics[height=9pt]{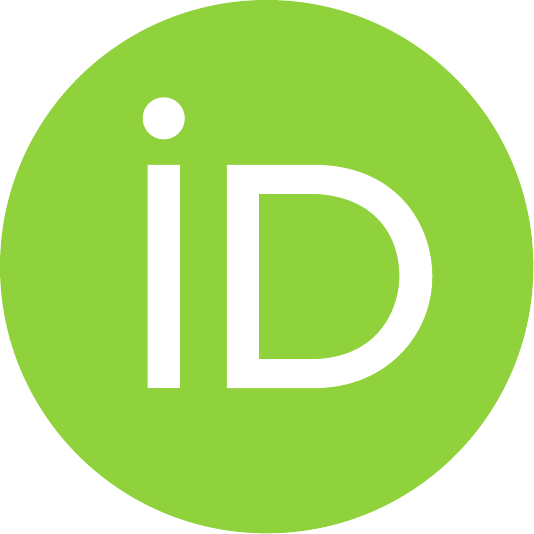}}}%
  \thanks{%
    Gernot Herbst is with University of Applied Sciences Zwickau, Germany (e-mail: \protect\href{mailto:gernot.herbst@fh-zwickau.de}{gernot.herbst@fh-zwickau.de}).%
  }
}

\markboth%
{Herbst: Transfer Function Analysis and Implementation of Active Disturbance Rejection Control}
{Herbst: Transfer Function Analysis and Implementation of Active Disturbance Rejection Control}

\maketitle

\begin{abstract}
To support the adoption of active disturbance rejection control (ADRC) in industrial practice, this article aims at improving both understanding and implementation of ADRC using traditional means, in particular via transfer functions and a frequency-domain view.
Firstly, to enable an immediate comparability with existing classical control solutions, a realizable transfer function implementation of continous-time linear ADRC is introduced.
Secondly, a frequency-domain analysis of ADRC components, performance, parameter sensitivity, and tuning method is performed.
Finally, an exact implementation of discrete-time ADRC using transfer functions is introduced for the first time, with special emphasis on practical aspects such as computational efficiency, low parameter footprint, and windup protection.
\end{abstract}

\begin{IEEEkeywords}
Active disturbance rejection control (ADRC), frequency-domain analysis, digital implementation.
\end{IEEEkeywords}


\section{Introduction}


\subsection{Motivation and related work}

Since its inception, active disturbance rejection control (ADRC) attracted widespread attention among both scholars and practitioners. Starting with Han's nonlinear controller \cite{Han:2009} and streamlined in Gao's linear variant \cite{Gao:2003}, it has found the way into numerous application domains. An overview  is given in \cite{Zheng:2010} and, more recently, in \cite{Zheng:2018}.

What are ADRC's main selling points? It is a serious---and probably the most popular---contender to overcome the ``theory versus practice hassle'' \cite{Huba:2019} that has been tackled also by other model-free controller approaches \cite{Fliess:2013}. The ``paradigm shift'' \cite{Gao:2006} is enabled by a combination of modern control elements with pragmatic, minimal plant modeling and control loop tuning efforts. At the same time, it can easily be equipped with controller features desirable in industrial practice \cite{Herbst:2016a}.

As a general purpose controller, ADRC competes with the ubiquitous PI and PID controllers. In many practical settings, these controllers are---and, which is a good thing, can be---tuned and implemented by people with expertise in the application domain, rather than dedicated control systems specialists. Therefore, to reach a wider audience with ADRC, we should adopt a suitable perspective and language to lower existing barriers for application experts. We believe this can be achieved with a frequency-domain view and implementation of linear ADRC.

In recent years, different studies have been tackling the connection of ADRC and transfer functions from various perspectives: \cite{Jin:2018} on the compensator properties of ADRC, \cite{Jin:2020} with a PID interpretation, \cite{Tan:2016} on the internal model control (IMC) representation of ADRC, \cite{Fu:2016} and \cite{Zhou:2019} on a transfer function representation and a generalization incorporating additional plant model information, and \cite{Zhou:2020} even on the opposite direction, implementing linear controllers with ADRC.

To achieve compatibility with existing traditional solutions in practical applications, simplifying ADRC to a one-degree of freedom (1DOF) error-based transfer function was proposed in \cite{Michalek:2016,Mandali:2020}, with an approximate discrete-time implementation being given as well \cite{Madonski:2019}. In contrast, retaining the original 2DOF characteristics of ADRC and providing exact continuous- and discrete-time transfer function representations are the guiding ideas pursued in this article.

Existing frequency-domain studies have put their focus on loop gain analysis \cite{Tian:2007,Xue:2013,Huang:2014,Zheng:2016}, disturbance behavior \cite{Zhang:2015,Alvarez:2017}, or the modified plant \cite{Zhang:2017:ICCAS}, but all without realizable transfer functions \cite{Zhao:2010,Huang:2013:CCC,Zhang:2014,Zhang:2017:BT,Zhang:2017:ASME,Jin:2018,Jin:2020}.
Departing from that, and to the best of the author's knowledge, this work presents the first study putting emphasis on realizable transfer functions, enabling both a comprehensive frequency-domain analysis of ADRC and its implementation at the same time.


\subsection{Structure and contributions of this article}

In order to establish the notation and some necessary design equations, a review of linear ADRC is presented in \refSec{sec:ADRC}. Afterwards, the main contributions of this article are presented in sections \ref{sec:TransferFunction}, \ref{sec:FrequencyDomain}, and \ref{sec:DiscreteTransferFunction}:
\begin{itemize}
\item
\emph{Implementing ADRC using transfer functions.}
\refSec{sec:TransferFunction} derives a transfer function representation of continuous-time ADRC, that---in contrast to existing works on this subject---has realizability in mind and seeks for similarity to traditional control loop setups. This will facilitate understanding and implementing ADRC using classical feedback controller and filter structures, with a special focus on first- and second-order ADRC as the contenders to PI and PID controllers.

\item
\emph{Understanding ADRC from a frequency-domain perspective.}
In \refSec{sec:FrequencyDomain} a detailed frequency-domain analysis is performed for continuous-time ADRC; including pole/zero analyses of its feedback controller transfer function and the influence of ADRC's tuning parameters on the closed-loop behavior. Again, a decidedly classical control systems perspective on ADRC is chosen in order to break down barriers hindering the adoption of state-space methods in real-world control systems.

\item
\emph{Efficiently implementing discrete-time ADRC using transfer functions.}
For discrete-time ADRC, an exact transfer function representation is derived in \refSec{sec:DiscreteTransferFunction} for the first time, enabling the implementation of discrete-time ADRC using digital filter structures. One of the main benefits is a reduction of the computational burden of the control law in comparison to the state-space form, which will increase the attractiveness of ADRC especially in resource-constrained embedded systems.
\end{itemize}


\section{Summary of continuous-time linear ADRC}
\label{sec:ADRC}


\subsection{Plant model}

As an introductory example we will, following related work \cite{Gao:2003,Herbst:2013}, consider a simple plant with $n$-th order low-pass behavior, output $y$, input $u$, and disturbance input $d$. Firstly, the input gain $b = b_0 + \Delta b$ is being split into a known ($b_0$) and unknown ($\Delta b$) part. Then, a generalized disturbance term $f$ is being introduced such that only a disturbed $n$-th order integrator chain remains of the original plant model:

\begin{align}
y^{(n)}(t)
&= \rlap{$\underbrace{ \phantom{-\sum_{i=0}^{n-1} a_i \cdot y^{(i)}(t) + e \cdot d(t) + \Delta b \cdot u(t)} }_{\text{generalized disturbance}\, f(t)}$} -\sum_{i=0}^{n-1} a_i \cdot y^{(i)}(t) + e \cdot d(t) + \overbrace{\Delta b \cdot u(t) + b_0 \cdot u(t)}^{= b \cdot u(t)}
\notag\\
&= f(t) + b_0 \cdot u(t)
\label{eqn:ADRC_Plant}
\end{align}

Considering any plant to be controlled as a disturbed integrator chain---regardless of the actual plant structure and parameters---is the gist of ADRC, and a distinct departure from model-based approaches \cite{Radke:2006}. Therefore, apart from the order of the system ($n$), $b_0$ remains as the only parameter that needs to be known for plant modeling in the context of ADRC. Due to its importance, $b_0$ is also known as the \emph{critical gain parameter} \cite{Madonski:2015}.


\subsection{Control law}

The control law of linear ADRC is based on three ingredients:
\begin{itemize}
\item
Inversion of the critical gain parameter, i.\,e.\ application of an inverse gain $\frac{1}{b_0}$ to the overall controller output,
\item
compensation of the generalized (total) disturbance using an estimated (extended) state $\hat{x}_{n + 1}$,
\item
a state-feedback controller (gains $k_1$ to $k_n$) for the remaining plant dynamics (normalized $n$-th integrator chain), based on a full-order observer, in order to achieve the desired closed-loop dynamics of the control loop.
\end{itemize}

\begin{figure}
    \centering%
    \includegraphics[width=\linewidth]{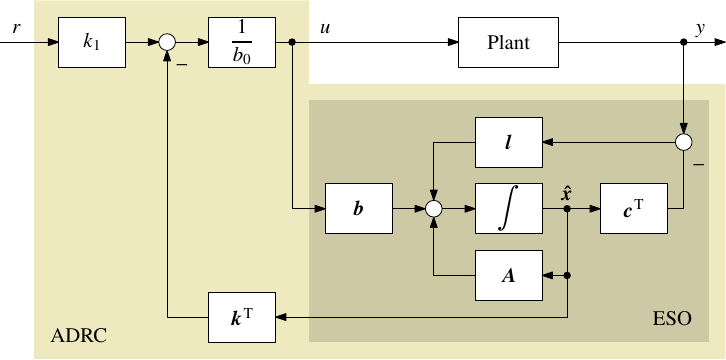}%
    \caption{Continuous-time state-space implementation of linear ADRC. The parameters for controller and extended state observer (ESO) are given in \refEq{eqn:ADRC_A_B_C} and \refEq{eqn:ADRC_K_L}.}
    \label{fig:ADRC_StateSpace}
\end{figure}

A visualization of the observer-based control law of linear ADRC is given in \refFig{fig:ADRC_StateSpace}.
With a reference signal $r(t)$ and controller output $u(t)$, the control law can be given as:

\begin{align}
u(t)
&= \frac{1}{b_0} \cdot \left( k_1 r(t) - \sum_{i = 1}^n k_i \hat{x}_i(t) - \hat{x}_{n + 1}(t) \right)
\notag\\
&= \frac{1}{b_0} \cdot \left( k_1 r(t) - \Vector{k}^\Transpose \Vector{\hat{x}}(t) \right)
\label{eqn:ADRC_ControlLaw}
\end{align}

A Luenberger observer of order $(n+1)$ is being set up in order to estimate the ($n$) states of the plant model (integrator chain), extended by the estimate $\hat{x}_{n + 1}$ of the generalized disturbance. In the context of ADRC, it is usually denoted as the \emph{extended state observer} \cite{Radke:2006,Miklosovic:2006} (ESO):
\begin{gather}
\Vector{\dot{\hat{x}}}(t)
= (\Matrix{A} - \Vector{l}\Vector{c}^\Transpose) \Vector{\hat{x}}(t) + \Vector{b} \Vector{u}(t) +  \Vector{l} y(t)
\label{eqn:ADRC_Observer}
\\
\text{with}\quad
\Vector{k}^\Transpose =
\begin{pmatrix}
k_1  &  \cdots  &  k_n  &  1
\end{pmatrix}
,\quad
\Vector{l} =
\begin{pmatrix}
l_1
\\
\vdots
\\
l_{n+1}
\end{pmatrix}
,
\label{eqn:ADRC_K_L}
\\
\Matrix{A} =
\begin{pmatrix}
\Matrix{0}^{n \times 1}  &  \Matrix{I}^{n \times n}
\\
0  &  \Matrix{0}^{1 \times n}
\end{pmatrix}
,\quad
\Vector{b} =
\begin{pmatrix}
\Matrix{0}^{(n-1) \times 1}
\\
b_0
\\
0
\end{pmatrix}
,\quad
\Vector{c}^\Transpose =
\begin{pmatrix}
1  &  \Matrix{0}^{1 \times n}
\end{pmatrix}
\label{eqn:ADRC_A_B_C}
\end{gather}

\subsection{Controller and observer design}

Throughout this article, the ``bandwidth parameterization'' approach \cite{Gao:2003} will be used, which is the most common method for designing the controller/observer in the context of linear ADRC. For the controller design we assume full rejection of the total (generalized) disturbance in \refEq{eqn:ADRC_Plant}. Therefore the characteristic polynomial of the closed control loop is being reduced to the following form (state feedback control of an $n$-th order integrator plant), and the gains $k_i$ can be obtained from parameterization with a desired closed-loop bandwidth $\omega_\mathrm{CL}$:%

\begin{align}
&
\det\left( s \Matrix{I} - \left(
\begin{pmatrix}
\Matrix{0}^{(n-1) \times 1}  &  \Matrix{I}^{(n-1) \times (n-1)}
\\
0  &  \Matrix{0}^{1 \times (n-1)}
\end{pmatrix}
\right.\right.
\notag\\
&
\quad\quad\quad\quad\left.\left.
-
\frac{1}{b_0} \cdot
\begin{pmatrix}
\Matrix{0}^{(n-1) \times 1}
\\
b_0
\end{pmatrix}
\cdot
\begin{pmatrix}
k_1  &  \cdots  &  k_n
\end{pmatrix}
\right) \right)
\notag\\
&= s^{n} + k_n s^{n-1} + \ldots + k_2 s + k_1
\stackrel{!}{=} \left( s + \omega_\mathrm{CL} \right)^n
\label{eqn:ADRC_Design_K_Approach}
\end{align}

In order to provide sufficiently fast state estimation and disturbance rejection, the closed-loop poles of the observer must be placed far enough left in the $s$-plane. We will follow the notation of \cite{Herbst:2013,Herbst:2016a} in bandwidth parameterization of the observer with common poles at $k_\mathrm{ESO} \cdot \omega_\mathrm{CL}$, with $k_\mathrm{ESO}$ being the relative factor (typically within the range $3 \ldots 10$) of the observer poles compared to the desired poles of the closed control loop:
\begin{align}
\det\left( s \Matrix{I} - \left( \Matrix{A} - \Vector{l} \Vector{c}^\Transpose \right) \right)
& = s^{n+1} + l_1 s^{n} + \ldots + l_{n} s + l_{n+1}
\notag\\
& \stackrel{!}{=} \left( s + k_\mathrm{ESO} \cdot \omega_\mathrm{CL} \right)^{n+1}
\label{eqn:ADRC_Design_L_Approach}
\end{align}

Solving \refEq{eqn:ADRC_Design_K_Approach} and \refEq{eqn:ADRC_Design_L_Approach} using binominal expansion of leads to the parameterization equations for $\Vector{k}^\Transpose$ and $\Vector{l}$. For $n = 1$ and $n = 2$, which are the most relevant cases in practice, the controller and observer gains can be found in \refTable{table:ControllerObserverGains}. The general terms are given in \refEq{eqn:ADRC_Design_K_Param} and \refEq{eqn:ADRC_Design_L_Param}.
\begin{equation}
k_i = \frac{n!}{(n-i+1)! \cdot (i-1)!} \cdot \omega_\mathrm{CL}^{n-i+1}
\quad \forall i = 1, ..., n
\label{eqn:ADRC_Design_K_Param}
\end{equation}
\begin{equation}
l_i = \frac{(n+1)!}{(n+1-i)! \cdot i!} \cdot \left( k_\mathrm{ESO} \cdot \omega_\mathrm{CL} \right)^i
\quad \forall i = 1, ..., n+1
\label{eqn:ADRC_Design_L_Param}
\end{equation}


\section{Transfer function representation}
\label{sec:TransferFunction}


\subsection{Approach}
\label{sec:TransferFunction_Approach}

Aim of this section is to derive a set of realizable transfer functions that implement ADRC such that a comparison with traditional controller and filter transfer functions becomes easily possible. This will also allow to analyze ADRC with traditional (frequency domain) means. We start with the closed-loop observer dynamics by putting the Laplace transform of \refEq{eqn:ADRC_ControlLaw} in \refEq{eqn:ADRC_Observer}:

\begin{equation}
\Vector{\hat{x}}(s)
= \left(s \Matrix{I} -  \Matrix{A}_\mathrm{CL} \right)^{-1}
\cdot \left( \frac{k_1}{b_0} \Vector{b} r(s) + \Vector{l} y(s) \right)
\label{eqn:Transfer_X}
\end{equation}
In \refEq{eqn:Transfer_X}, we abbreviated the system matrix of the closed-loop controller/observer dynamics as follows:
\begin{align}
\Matrix{A}_\mathrm{CL}
&=
\Matrix{A} - \Vector{l}\Vector{c}^\Transpose - \frac{1}{b_0} \Vector{b} \Vector{k}^\Transpose
\notag\\
&=
\begin{pmatrix}
\Matrix{0}^{n \times 1}  &  \Matrix{I}^{n \times n}
\\
0  &  \Matrix{0}^{1 \times n}
\end{pmatrix}
-
\begin{pmatrix}
\Vector{l}  &  \Matrix{0}^{(n+1) \times n}
\end{pmatrix}
-
\begin{pmatrix}
\Matrix{0}^{(n-1) \times (n+1)}
\\
\Vector{k}^\Transpose
\\
\Matrix{0}^{1 \times (n+1)}
\end{pmatrix}
\label{eqn:Transfer_ACL}
\end{align}
Putting \refEq{eqn:Transfer_X} back in the Laplace transform of \refEq{eqn:ADRC_ControlLaw} to eliminate $\Vector{\hat{x}}(s)$ yields the transfer function of the controller, which we immediately rewrite with a common denominator polynomial:
\begin{align}
u(s)
=&\ \frac{k_1}{b_0} r(s) - \frac{1}{b_0} \Vector{k}^\Transpose \cdot \left(s \Matrix{I} - \Matrix{A}_\mathrm{CL}\right)^{-1}
\cdot \left( \frac{k_1}{b_0} \Vector{b} r(s) + \Vector{l} y(s) \right)
\notag\\
=&\ \big( b_0 \det\left( s \Matrix{I} - \Matrix{A}_\mathrm{CL} \right) \big)^{-1} \cdot
\notag\\
&\ \Bigg[ \left( k_1 \det\left( s \Matrix{I} - \Matrix{A}_\mathrm{CL} \right) - \frac{k_1}{b_0} \Vector{k}^\Transpose \operatorname{adj}\left( s \Matrix{I} - \Matrix{A}_\mathrm{CL} \right) \Vector{b} \right) \cdot r(s)
\notag\\
&\ \ - \left(\Vector{k}^\Transpose \operatorname{adj}\left( s \Matrix{I} - \Matrix{A}_\mathrm{CL} \right) \Vector{l} \right) \cdot y(s) \Bigg]
\label{eqn:Transfer_U}
\end{align}

\begin{figure}
    \centering%
    \includegraphics[width=\linewidth]{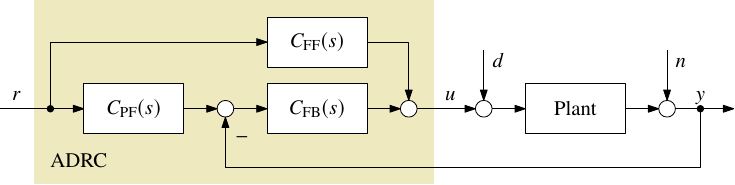}%
    \caption{Control loop with transfer function based ADRC implementation consisting of feedback controller $C_\mathrm{FB}(s)$, reference signal prefilter $C_\mathrm{PF}(s)$, and reference signal feedforward $C_\mathrm{FF}(s)$. Disturbances (at plant input, $d$, and plant output, $n$) are present in order to derive the gang-of-six transfer functions in \refSec{sec:FrequencyDomain_ClosedLoop}.}
    \label{fig:ADRC_TransferFunction}
\end{figure}

In this article, we propose a generic approach to represent continuous-time ADRC in frequency domain using three realizable transfer functions, namely a feedback controller $C_\mathrm{FB}(s)$, a reference signal prefilter $C_\mathrm{PF}(s)$, and a feedforward of the reference signal to the control signal via $C_\mathrm{FF}(s)$, cf.\ \refFig{fig:ADRC_TransferFunction}:
\begin{equation}
u(s) = C_\mathrm{FB}(s) \cdot \left[ C_\mathrm{PF}(s) \cdot r(s) - y(s) \right] + C_\mathrm{FF}(s) \cdot r(s)
\label{eqn:Transfer_U_Generic}
\end{equation}
The main benefits of the proposed transfer function approach are:
\begin{itemize}
\item
There is only one transfer function with an integrator component: the feedback controller $C_\mathrm{FB}(s)$, and its position in the control loop is exactly the same as in conventional control loops.
\item
By introducing the feedforward term $C_\mathrm{FF}(s)$, the reference signal prefilter transfer function $C_\mathrm{PF}(s)$ can be made realizable in the form of a conventional lead-lag filter. Otherwise its numerator polynomial order would exceed the denominator polynomial.
\end{itemize}

\subsection{Derivation of the transfer functions}
\label{sec:TransferFunction_Derivation}

The strictly proper feedback controller transfer function $C_\mathrm{FB}(s)$ can directly be obtained from \refEq{eqn:Transfer_U}:
\begin{equation}
C_\mathrm{FB}(s)
= \frac{ \Vector{k}^\Transpose \cdot \operatorname{adj}\left( s \Matrix{I} - \Matrix{A}_\mathrm{CL} \right) \cdot \Vector{l} }{ b_0 \cdot \det\left( s \Matrix{I} - \Matrix{A}_\mathrm{CL} \right) }
\label{eqn:Transfer_C_FB}
\end{equation}

Clearly $C_\mathrm{FB}(s)$ depends on the resolvent of $\Matrix{A}_\mathrm{CL}$. By inspection of \refEq{eqn:Transfer_ACL} we find $\Matrix{A}_\mathrm{CL}$ has rank $n$ (rightmost column is zero), assuming non-zero controller and observer gains. The denominator polynomial $\det\left( s \Matrix{I} - \Matrix{A}_\mathrm{CL} \right)$, which is of order $n+1$, will therefore introduce one integrator (pole at $s = 0$) to $C_\mathrm{FB}(s)$, see also \cite{Zhou:2019}.

By comparing \refEq{eqn:Transfer_U} and \refEq{eqn:Transfer_U_Generic} we obtain the following relation of $C_\mathrm{PF}(s)$ and $C_\mathrm{FF}(s)$ in combination with $C_\mathrm{FB}(s)$:
\begin{align}
&C_\mathrm{FB}(s) C_\mathrm{PF}(s) + C_\mathrm{FF}(s) =
\label{eqn:Transfer_C_FB_PF_FF}
\\
&\frac{ k_1 \cdot \det\left( s \Matrix{I} - \Matrix{A}_\mathrm{CL} \right) - \frac{k_1}{b_0} \cdot \Vector{k}^\Transpose \cdot \operatorname{adj}\left( s \Matrix{I} - \Matrix{A}_\mathrm{CL} \right) \cdot \Vector{b} }{ b_0 \cdot \det\left( s \Matrix{I} - \Matrix{A}_\mathrm{CL} \right) }
\notag
\end{align}

Since \refEq{eqn:Transfer_C_FB_PF_FF} is not strictly proper, $C_\mathrm{PF}(s)$ could not be made realizable without $C_\mathrm{FF}(s)$.%
\endnote[\textcolor{black!50}{\textdagger}]{%
    In this updated version of the preprint, the sentence after \refEq{eqn:Transfer_C_FB_PF_FF} includes a post-publication correction. The \href{https://doi.org/10.1007/s11768-021-00031-5}{official publication} incorrectly states ``$C_\mathrm{FB}(s)$ could not be made realizable without $C_\mathrm{FF}(s)$''. Instead, as mentioned before, it is $C_\mathrm{PF}(s)$---and not $C_\mathrm{FB}(s)$---which is affected by the realizability issue.
    The author would like to thank Rafal Madonski for reporting this issue.    
}
Therefore $C_\mathrm{FF}(s)$ will be chosen to carry the $s^{n+1}$ term from the numerator polynomial of \refEq{eqn:Transfer_C_FB_PF_FF}:
\begin{equation}
C_\mathrm{FF}(s)
= \frac{ k_1 \cdot s^{n+1} }{ b_0 \cdot \det\left( s \Matrix{I} - \Matrix{A}_\mathrm{CL} \right) }
\label{eqn:Transfer_C_FF}
\end{equation}

The actual order of numerator and denominator polynomial in $C_\mathrm{FF}(s)$ is only $n$ due to cancellation with the pole at origin introduced by $\det\left( s \Matrix{I} - \Matrix{A}_\mathrm{CL} \right)$, making $C_\mathrm{FF}(s)$ an $n$-th order high-pass filter. Now $C_\mathrm{PF}(s)$ becomes realizable in the form of a conventional $n$-th order lead-lag filter, and we can clearly see that the denominator will cancel the zeros of $C_\mathrm{FB}(s)$:
\begin{align}
\label{eqn:Transfer_C_PF}
&C_\mathrm{PF}(s) =
\\
&\frac{ k_1 \cdot \left( \det\left( s \Matrix{I} - \Matrix{A}_\mathrm{CL} \right) - s^{n+1} - \frac{1}{b_0} \cdot \Vector{k}^\Transpose \cdot \operatorname{adj}\left( s \Matrix{I} - \Matrix{A}_\mathrm{CL} \right) \cdot \Vector{b} \right) }{ \Vector{k}^\Transpose \cdot \operatorname{adj}\left( s \Matrix{I} - \Matrix{A}_\mathrm{CL} \right) \cdot \Vector{l} }
\notag
\end{align}

In summary, the three transfer functions required to implement a realizable form of $n$-th order ADRC can be formulated as follows:
\begin{align}
C_{\mathrm{FB},n}(s) &=
K_\mathrm{I} \cdot \frac{1}{s}
\cdot
\frac
{ 1 + \D\sum_{i = 1}^n \beta_i s^i }
{ 1 + \D\sum_{i = 1}^n \alpha_i s^i }
,
\\
C_{\mathrm{PF},n}(s) &=
\frac
{ 1 + \D\sum_{i = 1}^n \gamma s^i }
{ 1 + \D\sum_{i = 1}^n \beta_i s^i }
,
\\
C_{\mathrm{FF},n}(s) &=
\frac{ K_\mathrm{I} }{ l_{n+1} }
\cdot
\frac
{ s^n }
{ 1 + \D\sum_{i = 1}^n \alpha_i s^i }
.
\label{eqn:Transfer_C_FB_PF_FF_n}
\end{align}

For $n = 1$ and $n = 2$ (i.\,e.\ first- and second-order ADRC), the transfer functions as well as their parameters are given in \refTable{table:Transfer_C}, \refTable{table:TF_Parameters1}, and \refTable{table:TF_Parameters2}, respectively. Note that the coefficients in \refTable{table:TF_Parameters1} and \refTable{table:TF_Parameters2} are given both depending on the design parameters $k_\mathrm{ESO}$ and $\omega_\mathrm{CL}$ when using bandwidth parameterization, and in general terms using the controller/observer parameters $k$ and $l$. The latter allows to use alternative tuning approaches such as the recently proposed half-gain tuning for ADRC \cite{Herbst:2020a}.

\begin{table*}
\begin{center}
\caption{Continuous-time controller and observer gains using bandwidth parameterization. The tuning parameters are $\omega_\mathrm{CL}$ (desired closed-loop bandwidth) and $k_\mathrm{ESO}$ (observer bandwidth factor, i.\,e.\ relative position of the observer poles compared to the poles of the closed control loop).}
\label{table:ControllerObserverGains}
\begin{tabular*}{\linewidth}{@{\extracolsep\fill}rll@{\extracolsep\fill}}
    \toprule
    \textbf{}  &  \textbf{Controller gains}  &  \textbf{Observer gains}  \\
    \midrule
    \\[-0.9em]
    \textbf{First-order ADRC}
    &   $k_1 = \omega_\mathrm{CL}$
    &   $l_1 = 2 k_\mathrm{ESO} \omega_\mathrm{CL}$,\quad
        $l_2 = k_\mathrm{ESO}^2 \omega_\mathrm{CL}^2$
    \\[1.5em]
    \textbf{Second-order ADRC}
    &   $k_1 = \omega_\mathrm{CL}^2$,\quad
        $k_2 = 2 \omega_\mathrm{CL}$
    &   $l_1 = 3 k_\mathrm{ESO} \omega_\mathrm{CL}$,\quad
        $l_2 = 3 k_\mathrm{ESO}^2 \omega_\mathrm{CL}^2$,\quad
        $l_3 = k_\mathrm{ESO}^3 \omega_\mathrm{CL}^3$
    \\[0.9em]
    \bottomrule
\end{tabular*}
\end{center}
\end{table*}

\begin{table*}
\begin{center}
\caption{Continuous-time transfer function implementation of first- and second-order ADRC. The $\alpha$, $\beta$, $\gamma$ coefficients as well as the gains $K_\mathrm{I}$ and $K_\mathrm{I}/l_{n+1}$ for the first- ($n = 1$) and second order case ($n = 2$) can be obtained from \refTable{table:TF_Parameters1} and \refTable{table:TF_Parameters2}, respectively.}
\label{table:Transfer_C}
\begin{tabular*}{\linewidth}{@{\extracolsep\fill}rlll@{\extracolsep\fill}}
    \toprule
    \textbf{}  &  \textbf{Feedback controller}  &  \textbf{Reference signal prefilter}  &  \textbf{Reference signal feedforward}  \\
    \midrule
    \\[-0.9em]
    \textbf{First-order ADRC}
    &   $\D C_{\mathrm{FB},1}(s) = \frac{K_\mathrm{I}}{s} \cdot \frac{ 1 + \beta_1 s }{ 1 + \alpha_1 s }$
    &   $\D C_{\mathrm{PF},1}(s) = \frac{ 1 + \gamma_1 s }{ 1 + \beta_1 s }$
    &   $\D C_{\mathrm{FF},1}(s) = \frac{K_\mathrm{I}}{l_2} \cdot \frac{ s }{ 1 + \alpha_1 s }$
    \\[2.0em]
    \textbf{Second-order ADRC}
    &   $\D C_{\mathrm{FB},2}(s) = \frac{K_\mathrm{I}}{s} \cdot \frac{ 1 + \beta_1 s + \beta_2 s^2 }{ 1 + \alpha_1 s + \alpha_2 s^2 }$
    &   $\D C_{\mathrm{PF},2}(s) = \frac{ 1 + \gamma_1 s + \gamma_2 s^2 }{ 1 + \beta_1 s + \beta_2 s^2 }$
    &   $\D C_{\mathrm{FB},2}(s) = \frac{K_\mathrm{I}}{l_3} \cdot \frac{ s^2 }{ 1 + \alpha_1 s + \alpha_2 s^2 }$
    \\[1.5em]
    \bottomrule
\end{tabular*}
\end{center}
\end{table*}

\begin{table*}
\begin{center}
\caption{Continuous-time transfer function parameters for first-order ADRC. Bandwidth parameterization is based on $b_0$ (gain parameter of the plant model), $\omega_\mathrm{CL}$ (desired closed-loop bandwidth), and $k_\mathrm{ESO}$ (observer bandwidth factor).}
\label{table:TF_Parameters1}
\begin{tabular*}{\linewidth}{@{\extracolsep\fill}rll@{\extracolsep\fill}}
    \toprule
    \textbf{Parameter}  &  \textbf{General terms}  &  \textbf{Bandwidth parameterization}  \\
    \midrule
    \\[-0.9em]
    $K_\mathrm{I}$  &
    $\D\frac{1}{b_0} \cdot \frac{ k_1 l_2 }{ k_1 + l_1 }$  &
    $\D\frac{1}{b_0} \cdot \frac{ k_\mathrm{ESO}^2 \omega_\mathrm{CL}^2 }{ 1 + 2 k_\mathrm{ESO} }$
    \\[1.2em]
    $\D\frac{K_\mathrm{I}}{l_2}$  &
    $\D\frac{1}{b_0} \cdot \frac{ k_1 }{ k_1 + l_1 }$  &
    $\D\frac{1}{b_0} \cdot \frac{ 1 }{ 1 + 2 k_\mathrm{ESO} }$
    \\[1.2em]
    $\alpha_1$  &
    $\D\frac{ 1 }{ k_1 + l_1 }$  &
    $\D\frac{ 1 }{ \omega_\mathrm{CL} \cdot \left( 1 + 2 k_\mathrm{ESO} \right) }$
    \\[1.2em]
    $\beta_1$  &
    $\D\left( \frac{l_1}{l_2} + \frac{ 1 }{ k_1 } \right)$  &
    $\D\frac{ 2 + k_\mathrm{ESO} }{ k_\mathrm{ESO} \omega_\mathrm{CL} }$
    \\[1.2em]
    $\gamma_1$  &
    $\D\frac{ l_1 }{ l_2 }$  &
    $\D\frac{ 2 }{ k_\mathrm{ESO} \omega_\mathrm{CL} }$
    \\[0.9em]
    \bottomrule
\end{tabular*}
\end{center}
\end{table*}

\begin{table*}
\begin{center}
\caption{Continuous-time transfer function parameters for second-order ADRC. Bandwidth parameterization is based on $b_0$ (gain parameter of the plant model), $\omega_\mathrm{CL}$ (desired closed-loop bandwidth), and $k_\mathrm{ESO}$ (observer bandwidth factor).}
\label{table:TF_Parameters2}
\begin{tabular*}{\linewidth}{@{\extracolsep\fill}rll@{\extracolsep\fill}}
    \toprule
    \textbf{Parameter}  &  \textbf{General terms}  &  \textbf{Bandwidth parameterization}  \\
    \midrule
    \\[-0.9em]
    $K_\mathrm{I}$  &
    $\D\frac{1}{b_0} \cdot \frac{ k_1 l_3 }{ k_1 + k_2 l_1 + l_2 }$  &
    $\D\frac{1}{b_0} \cdot \frac{ k_\mathrm{ESO}^3 \omega_\mathrm{CL}^3 }{ 1 + 6 k_\mathrm{ESO} + 3 k_\mathrm{ESO}^2 }$
    \\[1.2em]
    $\D\frac{K_\mathrm{I}}{l_3}$  &
    $\D\frac{1}{b_0} \cdot \frac{ k_1 }{ k_1 + k_2 l_1 + l_2 }$  &
    $\D\frac{1}{b_0} \cdot \frac{ 1 }{ 1 + 6 k_\mathrm{ESO} + 3 k_\mathrm{ESO}^2 }$
    \\[1.2em]
    $\alpha_1$  &
    $\D\frac{ k_2 + l_1 }{ k_1 + k_2 l_1 + l_2 }$  &
    $\D\frac{ 2 + 3 k_\mathrm{ESO} }{ \omega_\mathrm{CL} \cdot \left( 1 + 6 k_\mathrm{ESO} + 3 k_\mathrm{ESO}^2 \right) }$
    \\[1.2em]
    $\alpha_2$  &
    $\D\frac{ 1 }{ k_1 + k_2 l_1 + l_2 }$  &
    $\D\frac{ 1 }{ \omega_\mathrm{CL}^2 \cdot \left( 1 + 6 k_\mathrm{ESO} + 3 k_\mathrm{ESO}^2 \right) }$
    \\[1.2em]
    $\beta_1$  &
    $\D\left( \frac{l_2}{l_3} + \D\frac{ k_2 }{ k_1 } \right)$  &
    $\D\frac{ 1 }{ \omega_\mathrm{CL} } \cdot \left( \frac{3}{k_\mathrm{ESO}} + 2 \right)$
    \\[1.2em]
    $\beta_2$  &
    $\D\left( \frac{l_1}{l_3} + \frac{ k_2 }{ k_1 } \cdot \frac{l_2}{l_3} + \frac{ 1 }{ k_1 } \right)$  &
    $\D\frac{ 1 }{ \omega_\mathrm{CL}^2 } \cdot \left( \frac{3}{k_\mathrm{ESO}^2} + \frac{ 6 }{ k_\mathrm{ESO} } + 1 \right)$
    \\[1.2em]
    $\gamma_1$  &
    $\D\frac{ l_2 }{ l_3 }$  &
    $\D\frac{ 3 }{ k_\mathrm{ESO} \omega_\mathrm{CL} }$
    \\[1.2em]
    $\gamma_2$  &
    $\D\frac{ l_1 }{ l_3 }$  &
    $\D\frac{ 3 }{ k_\mathrm{ESO}^2 \omega_\mathrm{CL}^2 }$
    \\[1.2em]
    \bottomrule
\end{tabular*}
\end{center}
\end{table*}


\section{Frequency-domain analysis}
\label{sec:FrequencyDomain}


\subsection{Plant model: The critical gain parameter in frequency domain}
\label{sec:FrequencyDomain_b0}

For a frequency-domain interpretation of the important plant model parameter $b_0$ (also known as the \emph{critical gain parameter} \cite{Madonski:2015}), we will start by rewriting \refEq{eqn:ADRC_Plant} as a frequency-domain transfer function:
\begin{align}
P(\mathrm{j}\omega) &= \frac{y(\mathrm{j}\omega)}{u(\mathrm{j}\omega)}
\notag\\
&= \frac{ b_0 }{ (\mathrm{j}\omega)^n + a_{n-1} \cdot (\mathrm{j}\omega)^{n-1} + \ldots + a_1 \cdot \mathrm{j}\omega + a_0 }
\label{eqn:ADRC_Plant_jw}
\end{align}

For increasing angular frequencies $\omega$ the magnitude $\left| P(\mathrm{j}\omega) \right|$ will be dominated by the highest-order term in its denominator:
\begin{equation}
\lim_{\omega \to \infty} \left| P(\mathrm{j}\omega) \right|
\approx \left| \frac{ b_0 }{ (\mathrm{j}\omega)^n } \right|
= \frac{ b_0 }{ \omega^n }
\end{equation}

Solving this high-frequency approximation for its crossover frequency $\omega_\mathrm{X}$ yields the relation between $\omega_\mathrm{X}$ and $b_0$:

\begin{equation}
b_0 = \omega_\mathrm{X}^n
\quad\text{or}\quad
\omega_\mathrm{X} = \sqrt[^n]{b_0}
\end{equation}

\begin{figure*}
    \centering%
    \includegraphics{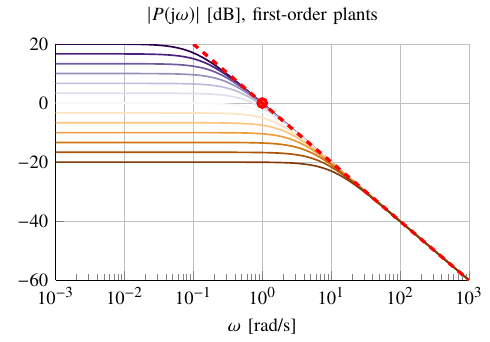}%
    \includegraphics{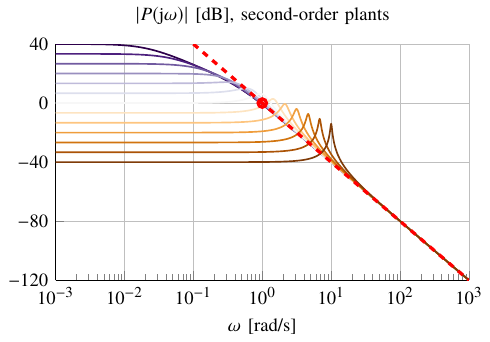}%
    \caption{Frequency-domain illustration of the critical gain parameter $b_0$ for first- (left-hand side) and second-order (right-hand side diagram) plant variations with the same value of $b_0$ (in these examples: $b_0 = 1$). The critical gain parameter can be obtained from  magnitude plots of the plant transfer function by extending the straight-line approximation (red dashed line) of the $-20$\,dB/decade (first-order case) or $-40$\,dB/decade segment (second-order case) in order to find the $0$\ dB crossover frequency (encircled in the diagrams). The crossover angular frequency amounts to $b_0$ in the first-order case and $\sqrt{b_0}$ in the second-order case.}
    \label{fig:FrequencyDomain_Plant_b0}
\end{figure*}

For $n$-th order plants that can be represented by \refEq{eqn:ADRC_Plant_jw}, this means that the corresponding $b_0$ parameter can be found from the plant's Bode (magnitude) plot by extending the straight-line approximation of the $-n \cdot 20\,\mathrm{dB}$/decade segment in order to find its crossover frequency $\omega_\mathrm{X}$. A visualization of this relation for first- and second-order cases is given in \refFig{fig:FrequencyDomain_Plant_b0}.

\subsection{Analysis and comparison of the feedback controller transfer function}
\label{sec:FrequencyDomain_Controller}


\subsubsection{Feedback controller of first-order ADRC}

Let us consider the feedback controller transfer function from \refTable{table:Transfer_C} using bandwidth parameterization as given in \refTable{table:TF_Parameters1}:
\begin{equation}
C_\mathrm{FB,1}(s) =
\underbrace{ \D\frac{1}{b_0} \cdot \frac{ k_\mathrm{ESO}^2 \omega_\mathrm{CL}^2 }{ 1 + 2 k_\mathrm{ESO} }  }_{ K_\mathrm{I} } \cdot \frac{1}{s}
\cdot
\frac
{ 1 + \overbrace{ \frac{ 2 + k_\mathrm{ESO} }{ k_\mathrm{ESO} \omega_\mathrm{CL} } }^{ \beta_1 = 1/\omega_\mathrm{Z} } \cdot s }
{ 1 + \underbrace{ \frac{ 1 }{ \omega_\mathrm{CL} \cdot (1 + 2 k_\mathrm{ESO}) } }_{ \alpha_1 = 1/\omega_\mathrm{P} } \cdot s }
\label{eqn:transfer_C_FB1_bw}
\end{equation}

This controller structure (PI controller with first-order noise filter, or interpretation as integrator with lead/lag filter) is a common choice for example in power electronics applications, known as ``Type 2'' controller \cite{Venable:1983}. It is also in line with the recommendation of H\"{a}gglund to add a first-order noise filter to PI controllers \cite{Haegglund:2012}.

\begin{figure*}
    \centering%
    \includegraphics{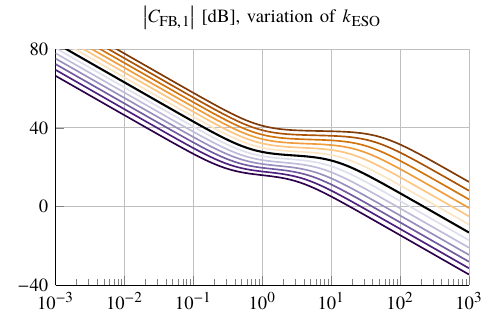}%
    \includegraphics{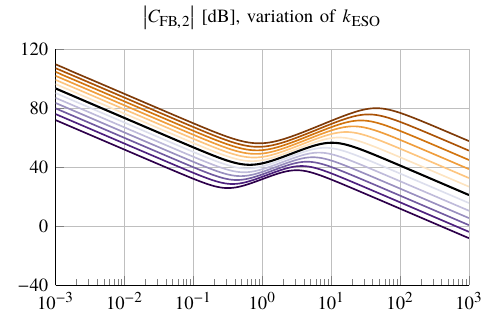}%
    \\%
    \includegraphics{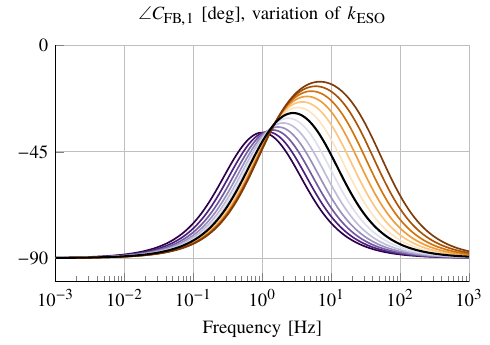}%
    \includegraphics{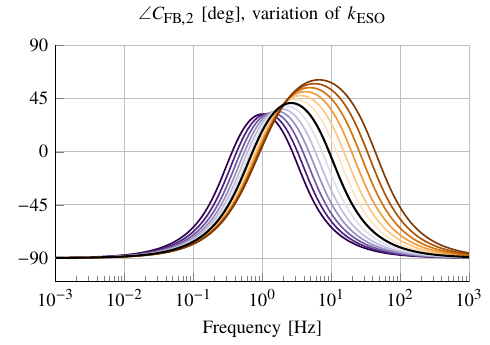}%
    \caption{Bode plots of the feedback controller $C_\mathrm{FB}$ of first- and second-order ADRC with variation of $k_\mathrm{ESO}$ using bandwidth parameterization. The exponentially increasing values of $k_\mathrm{ESO}$ are ranging from $k_\mathrm{ESO} = 1$ (purple) through $k_\mathrm{ESO} = 5$ (black) to $k_\mathrm{ESO} = 25$ (orange). The remaining tuning parameters are fixed at $b_0 = 1$ and $\omega_\mathrm{CL} = 2\piup$ (i.\,e.\ $1\,\mathrm{Hz}$).}
    \label{fig:FrequencyDomain_C_FB}
\end{figure*}

A frequency-domain visualization of the impact of the tuning parameter $k_\mathrm{ESO}$ on the feedback controller transfer function, similar to the analysis performed in \cite{Zhang:2014}, is given in \refFig{fig:FrequencyDomain_C_FB}, while the effect of $b_0$ and $\omega_\mathrm{CL}$ on $K_\mathrm{I}$ and the pole/zero frequencies can be easily grasped from \refEq{eqn:transfer_C_FB1_bw}. For high observer gains (increasing values of $k_\mathrm{ESO}$), the low-pass filter cut-off frequency moves to infinity, therefore the feedback controller transfer function \refEq{eqn:transfer_C_FB1_bw} converges to a standard PI controller with a zero at the desired closed-loop bandwidth:
\begin{equation}
\lim_{k_\mathrm{ESO} \to \infty} \omega_\mathrm{Z}
= \lim_{k_\mathrm{ESO} \to \infty} \frac{ k_\mathrm{ESO} \omega_\mathrm{CL} }{ 2 + k_\mathrm{ESO} }
= \omega_\mathrm{CL}
\end{equation}

\subsubsection{Feedback controller of second-order ADRC}

The feedback controller transfer function $C_\mathrm{FB,2}$ from \refTable{table:Transfer_C} can be interpreted as an integrator with second-order lead/lag filter, a structure known as ``Type 3'' controller in power electronics applications \cite{Venable:1983}. Alternatively, it can be viewed as a PID controller with a second-order low-pass noise filter, as also pointed out in \cite{Jin:2020}.
As a side note, this structure is recommended for PID by H\"{a}gglund \cite{Haegglund:2012}.

The poles of $C_\mathrm{FB,2}(s)$ consist of an integrator (pole at origin) and the poles defined by the roots of $\left( 1 + \alpha_1 s + \alpha_2 s^2 \right)$. Rewriting the latter part of the denominator polynomial using bandwidth parameterization as given in \refTable{table:TF_Parameters2} leads to \refEq{eqn:transfer_C_FB2_denom}.

\begin{align}
s^2 + \frac{\alpha_1}{\alpha_2} \cdot s + \frac{1}{\alpha_2}
=\ &s^2 + \left( 2 \omega_\mathrm{CL} + 3 k_\mathrm{ESO} \omega_\mathrm{CL} \right) \cdot s
\notag\\
&+ \left( \omega_\mathrm{CL}^2 + 6 k_\mathrm{ESO} \omega_\mathrm{CL}^2 + 3 k_\mathrm{ESO}^2 \omega_\mathrm{CL}^2 \right)
\label{eqn:transfer_C_FB2_denom}
\end{align}

Solving for the the roots of \refEq{eqn:transfer_C_FB2_denom} gives the conjugate complex poles $s_\mathrm{P1/2}$ of $C_\mathrm{FB,2}$:

\begin{equation}
s_\mathrm{P1/2} = \omega_\mathrm{CL} \cdot \left[ -\left( 1 + \frac{3}{2} k_\mathrm{ESO} \right) \pm \mathrm{j} \sqrt{ 3 k_\mathrm{ESO} + \frac{3}{4} k_\mathrm{ESO}^2 } \right]
\label{eqn:transfer_C_FB2_poles}
\end{equation}

Similarly one can obtain the zeros of $C_\mathrm{FB,2}$ by rewriting the numerator polynomial $\left( 1 + \beta_1 s + \beta_2 s^2 \right)$ using bandwidth parameterization:
\begin{align}
s^2 + \frac{\beta_1}{\beta_2} \cdot s + \frac{1}{\beta_2}
=\ & s^2 + \frac{ \left( 3 + 2 k_\mathrm{ESO} \right) k_\mathrm{ESO} \omega_\mathrm{CL} }{ 3 + 6 k_\mathrm{ESO} + k_\mathrm{ESO}^2 } \cdot s
\notag\\
&+ \frac{ k_\mathrm{ESO}^2 \omega_\mathrm{CL}^2 }{ 3 + 6 k_\mathrm{ESO} + k_\mathrm{ESO}^2 }
\label{eqn:transfer_C_FB2_num}
\end{align}
Solving for the the roots of \refEq{eqn:transfer_C_FB2_num} gives the conjugate complex zeros $s_\mathrm{Z1/2}$ of $C_\mathrm{FB,2}$:
\begin{equation}
s_\mathrm{Z1/2} = \frac{ k_\mathrm{ESO} \omega_\mathrm{CL} }{ 3 + 6 k_\mathrm{ESO} + k_\mathrm{ESO}^2 } \cdot \left[ -\left( \frac{3}{2} + k_\mathrm{ESO} \right) \pm \mathrm{j} \sqrt{ \frac{3}{4} + 3 k_\mathrm{ESO} } \right]
\label{eqn:transfer_C_FB2_zeros}
\end{equation}

For $k_\mathrm{ESO} \to \infty$, the poles $s_\mathrm{P1/2}$ remain conjugate complex with a rather fixed damping ratio starting at $D \approx 0.791$ for $k_\mathrm{ESO} = 1$ and relatively quickly converging to $D = \sqrt{3/4} \approx 0.866$. Interestingly, this fits in the design constraints (damping $D \in [1/\sqrt{2}, 1]$) given by Larsson for the design of second-order noise filters for PID controllers \cite{Larsson:2011}.

The real part of $s_\mathrm{P1/2}$ moves towards minus infinity with $k_\mathrm{ESO} \to \infty$, while \refEq{eqn:transfer_C_FB2_zeros} converges to a pair of real zeros with common location $s_\mathrm{Z1/2} = -\omega_\mathrm{CL}$. Therefore one can conclude that with increasing values of $k_\mathrm{ESO}$ the feedback controller part of second-order ADRC converges to a PID controller.
Note that this relation was independently reported very recently in \cite{Jin:2020}, as well.

The impact of $k_\mathrm{ESO}$ on $C_\mathrm{FB,2}(\mathrm{j}\omega)$ is illustrated in \refFig{fig:FrequencyDomain_C_FB}.
A report on the resulting phase margins can be found in \cite{Jin:2018}.


\subsection{Influence of tuning parameters on closed-loop behavior}
\label{sec:FrequencyDomain_ClosedLoop}

ADRC is a two-degree-of-freedoms controller, and as such, a control loop with ADRC is only being fully characterized by six transfer functions known as the ``gang of six'' \cite{AstromMurray:2008}. They relate the inputs $r$, $d$, $n$ of a control loop as depicted in \refFig{fig:ADRC_TransferFunction} to the signals $y$ and $u$, cf.\ \refEq{eqn:GangOfSix}. For the realizable transfer function representation of continuous-time ADRC derived in \refSec{sec:TransferFunction}, the gang-of-six transfer functions are given in \refTable{table:TF_GangOfSix}.
\begin{equation}
\begin{pmatrix}
y(s)
\\
u(s)
\end{pmatrix}
=
\begin{pmatrix}
G_{yr}(s)  &  G_{yd}(s)  &  G_{yn}(s)
\\
G_{ur}(s)  &  G_{ud}(s)  &  G_{un}(s)
\end{pmatrix}
\cdot
\begin{pmatrix}
r(s)
\\
d(s)
\\
n(s)
\end{pmatrix}
\label{eqn:GangOfSix}
\end{equation}

\begin{figure*}
    \centering%
    \includegraphics[]{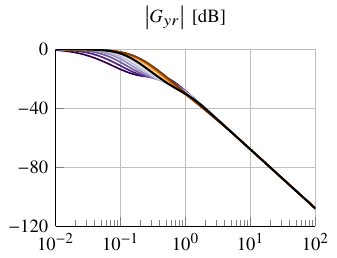}%
    \includegraphics[]{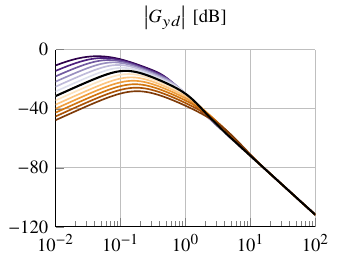}%
    \includegraphics[]{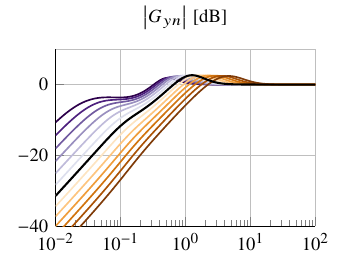}%
    \\%
    \includegraphics[]{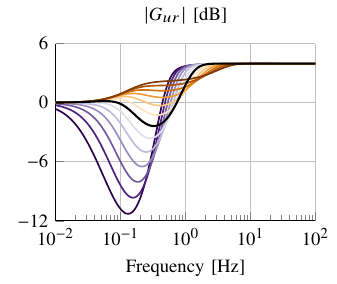}%
    \includegraphics[]{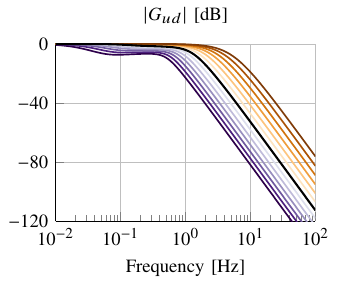}%
    \includegraphics[]{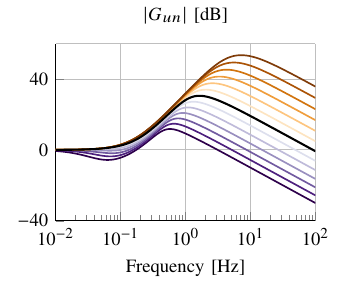}%
    \caption{Sensitivity of the gang-of-six transfer functions to variations of $k_\mathrm{ESO}$ for second-order ADRC. Normalized plant in this example: $P(s) = \frac{1}{1 + 2s + s^2}$. Fixed tuning parameters are $b_0 = 1$ and $\omega_\mathrm{CL} = 0.4\piup$ (i.\,e.\ $0.2\,\mathrm{Hz}$). The exponentially increasing values of $k_\mathrm{ESO}$ are ranging from $k_\mathrm{ESO} = 1$ (purple) through $k_\mathrm{ESO} = 5$ (black) to $k_\mathrm{ESO} = 25$ (orange).}
    \label{fig:FrequencyDomain_GangOfSix_ADRC2_kESO}
\end{figure*}

\begin{figure*}
    \centering%
    \includegraphics[]{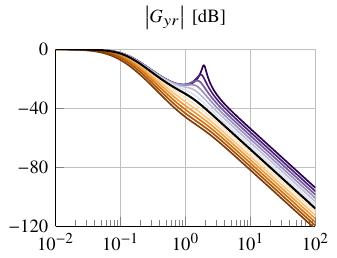}%
    \includegraphics[]{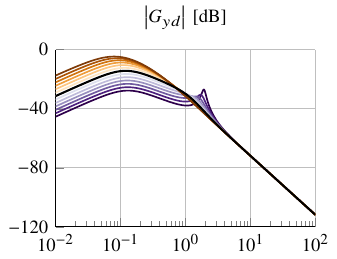}%
    \includegraphics[]{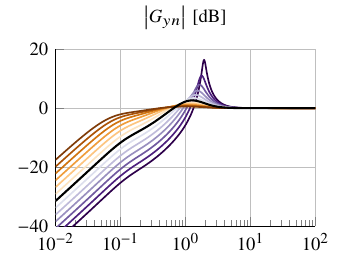}%
    \\%
    \includegraphics[]{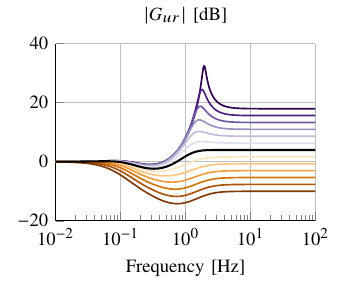}%
    \includegraphics[]{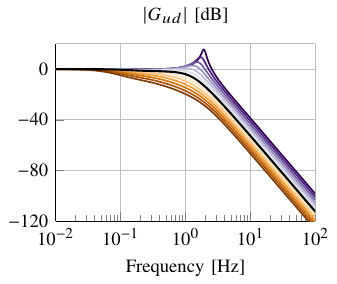}%
    \includegraphics[]{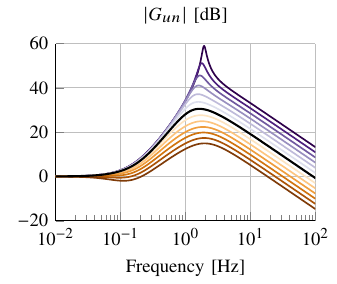}%
    \caption{Sensitivity of the gang-of-six transfer functions to variations of $b_0$ for second-order ADRC. Normalized plant in this example: $P(s) = \frac{1}{1 + 2s + s^2}$. Fixed tuning parameters are $\omega_\mathrm{CL} = 0.4\piup$ (i.\,e.\ $0.2\,\mathrm{Hz}$) and $k_\mathrm{ESO} = 5$. The exponentially increasing values of $b_0$ are ranging from $b_0 = 0.2$ (purple) through $b_0 = 1$ (black, nominal value) to $b_0 = 5$ (orange).}
    \label{fig:FrequencyDomain_GangOfSix_ADRC2_b0}
\end{figure*}

\begin{figure*}
    \centering%
    \includegraphics[]{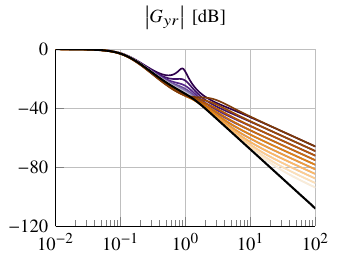}%
    \includegraphics[]{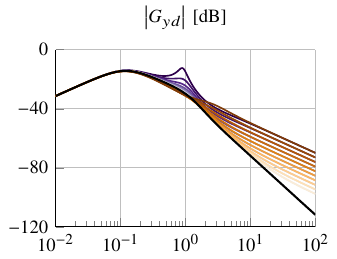}%
    \includegraphics[]{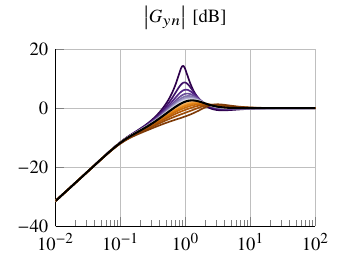}%
    \\%
    \includegraphics[]{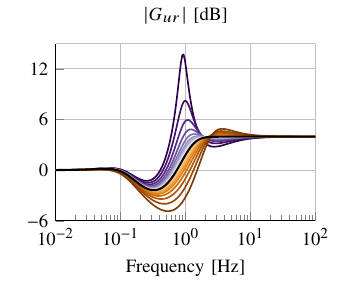}%
    \includegraphics[]{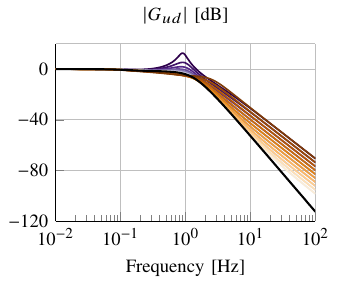}%
    \includegraphics[]{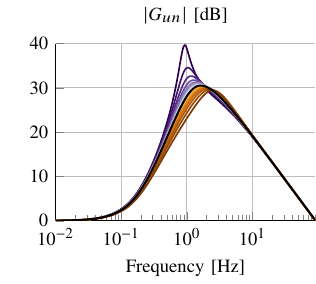}%
    \caption{Sensitivity of the gang-of-six transfer functions to an additional plant zero for second-order ADRC, covering both right half plane (RHP) and left half plane plane zeros. Normalized plant in this example (without the additional zero): $P(s) = \frac{1}{1 + 2s + s^2}$. Fixed tuning parameters are $b_0 = 1$, $\omega_\mathrm{CL} = 0.4\piup$ (i.\,e.\ $0.2\,\mathrm{Hz}$), $k_\mathrm{ESO} = 5$. The time constant of the additional zero ranges from $-0.2\,\mathrm{s}$ (RHP zeros, purple) to $0.2\,\mathrm{s}$ (orange).}
    \label{fig:FrequencyDomain_GangOfSix_ADRC2_Zero}
\end{figure*}

\begin{table*}
\begin{center}
\caption{Gang-of-six transfer functions for a closed control loop with plant $P(s)$ and continuous-time ADRC in a transfer function implementation with feedback controller $C_\mathrm{FB}(s)$, reference signal prefilter $C_\mathrm{PF}(s)$, and feedforward term $C_\mathrm{FF}(s)$.}
\label{table:TF_GangOfSix}
\begin{tabular*}{\linewidth}{@{\extracolsep\fill}rlll@{\extracolsep\fill}}
    \toprule
    \textbf{}  &  \textbf{From: $r$ (reference signal)}  &  \textbf{From: $d$ (disturbance)}  &  \textbf{From: $n$ (noise)}  \\
    \midrule
    \\[-0.9em]
    \textbf{To: $y$}
    &   $\D G_{yr}(s) = \frac{P(s) \cdot \left( C_\mathrm{FF}(s) + C_\mathrm{FB}(s) C_\mathrm{PF}(s) \right)}{1 + P(s) C_\mathrm{FB}(s)}$
    &   $\D G_{yd}(s) = \frac{P(s)}{1 + P(s) C_\mathrm{FB}(s)}$
    &   $\D G_{yn}(s) = \frac{1}{1 + P(s) C_\mathrm{FB}(s)}$
    \\[2.0em]
    \textbf{To: $u$}
    &   $\D G_{ur}(s) = \frac{C_\mathrm{FF}(s) + C_\mathrm{FB}(s) C_\mathrm{PF}(s)}{1 + P(s) C_\mathrm{FB}(s)}$
    &   $\D G_{ud}(s) = \frac{-P(s) C_\mathrm{FB}(s)}{1 + P(s) C_\mathrm{FB}(s)}$
    &   $\D G_{un}(s) = \frac{-C_\mathrm{FB}(s)}{1 + P(s) C_\mathrm{FB}(s)}$
    \\[0.9em]
    \bottomrule
\end{tabular*}
\end{center}
\end{table*}

\begin{table*}
\begin{center}
\caption{Discrete-time controller and observer gains using bandwidth parameterization. Note that the controller gains do not change compared to the continuous-time version, cf.\ \refTable{table:ControllerObserverGains}; only the observer dynamics have to be discretized \cite{Herbst:2013}. The tuning parameters are $T$ (sample time of the discretized implementation) and $z_\mathrm{ESO} = \mathrm{e}^{- k_\mathrm{ESO} \omega_\mathrm{CL} T}$ (observer pole locations in $z$-domain, based on the desired closed-loop bandwidth $\omega_\mathrm{CL}$ and the observer bandwidth factor $k_\mathrm{ESO}$).}
\label{table:ControllerObserverGains_Discrete}
\begin{tabular*}{\linewidth}{@{\extracolsep\fill}rll@{\extracolsep\fill}}
    \toprule
    \textbf{}  &  \textbf{Controller gains}  &  \textbf{Observer gains}  \\
    \midrule
    \\[-0.9em]
    \textbf{First-order ADRC}
    &   $k_1 = \omega_\mathrm{CL}$
    &   $\D l_1 = 1 - z_\mathrm{ESO}^2$,\quad
        $\D l_2 = \frac{1}{T} \cdot \left(1 - z_\mathrm{ESO}\right)^2$
    \\[1.5em]
    \textbf{Second-order ADRC}
    &   $k_1 = \omega_\mathrm{CL}^2$,\quad
        $k_2 = 2 \omega_\mathrm{CL}$
    &   $\D l_1 = 1 - z_\mathrm{ESO}^3$,\quad
        $\D l_2 = \frac{3}{2 T} \cdot \left(1 - z_\mathrm{ESO}\right)^2 \cdot \left(1 + z_\mathrm{ESO}\right)$,\quad
        $\D l_3 = \frac{1}{T^2} \cdot \left(1 - z_\mathrm{ESO}\right)^3$
    \\[0.9em]
    \bottomrule
\end{tabular*}
\end{center}
\end{table*}

With the full picture the gang-of-six transfer functions provide, it is possible to gain deeper insights into the trade-offs that must be resolved when tuning the parameters of ADRC. When the desired closed-loop bandwidth $\omega_\mathrm{CL}$ is chosen (depending on the plant dynamics as well as the requirements of the application), there are two questions remaining: How to choose the relative bandwidth $k_\mathrm{ESO}$ of the observer? What happens if the plant model parameter $b_0$ is above or below the actual value of the plant, i.\,e., is it better to over- or to underestimate $b_0$?

\subsubsection{Influence of \texorpdfstring{$k_\mathrm{ESO}$}{kESO} variations}
\label{sec:FrequencyDomain_Influence_kESO}

A typical choice \cite{Herbst:2013} of $k_\mathrm{ESO}$ would be in the range of $3 \ldots 10$. To provide a broader view, an extended interval of values ranging from $k_\mathrm{ESO} = 1$ to $k_\mathrm{ESO} = 25$ has been selected in the visual gang-of-six sensitivity analysis performed in \refFig{fig:FrequencyDomain_GangOfSix_ADRC2_kESO} for a normalized plant $P(s) = \frac{1}{1 + 2s + s^2}$.

Clearly, higher values of $k_\mathrm{ESO}$ (i.\,e.\ faster observers) help rejecting low-frequency disturbances at plant input ($d$) and output ($n$), and maintain the desired dynamics between reference signal ($r$) and controlled variable ($y$). On the other hand, a price is to pay in increasing high-frequency gains of transfer functions from input and output disturbances to the control action ($u$). Especially from $G_{un}$ it can be seen that the impact of high-frequency measurement noise on the control action will be a limiting factor (upper bound) for selecting $k_\mathrm{ESO}$.

When tuning $k_\mathrm{ESO}$ it is advisable to start with a single-digit value. From \refFig{fig:FrequencyDomain_GangOfSix_ADRC2_kESO}, the following practical guidelines can be inferred:
\begin{itemize}
\item
The low-frequency disturbance rejection capabilities can be improved by increasing $k_\mathrm{ESO}$, as long as the impact of measurement noise on the control signal is still acceptable. Note from $G_{un}$ in \refFig{fig:FrequencyDomain_GangOfSix_ADRC2_kESO} that boosting  $k_\mathrm{ESO}$ by a factor of five comes with an increased high-frequency noise sensitivity gain of almost $+40\,\mathrm{dB}$ in this example!

\item
The desired closed-loop dynamics $\omega_\mathrm{CL}$ should be attainable with a value of $k_\mathrm{ESO} \gtrapprox 5$. If this is not the case, $b_0$ might be off, cf.\ \refSec{sec:FrequencyDomain_Influence_b0}.
\end{itemize}

\subsubsection{Influence of \texorpdfstring{$b_0$}{b0} variations}
\label{sec:FrequencyDomain_Influence_b0}

\refFigBegin{fig:FrequencyDomain_GangOfSix_ADRC2_b0} shows the effect of under- and overestimating $b_0$ on the closed-loop dynamics using the same exemplary plant, both by a factor of up to five. Overestimating $b_0$ will lead to a less aggressively tuned controller with lower noise sensitivity, of course being also less effective in compensating low-frequency disturbances at the plant input or output. A compromise has to be found for $b_0$ since, on the other hand, underestimating $b_0$---while providing better disturbance rejection at the cost of increased high-frequency noise sensitivity---can induce pronounced medium-frequency oscillations in both control signal ($u$) and plant output ($y$), which will eventually lead to instability.
Note that the frequency of the oscillations that will occur when underestimating $b_0$ can be read from any of the gang-of-six plots in \refFig{fig:FrequencyDomain_GangOfSix_ADRC2_b0}.

As a guideline when choosing $b_0$ in cases its value cannot exactly be inferred from a plant model, the following procedure is recommended in light of the insights provided by the gang-of-six analysis:
\begin{itemize}
\item
Start with overestimating $b_0$ to obtain a stable loop. A too large value of $b_0$ can be identified by a reduced bandwidth of the closed loop compared to the design value $\omega_\mathrm{CL}$ (cf.\ orange plots of $G_{yr}$ in \refFig{fig:FrequencyDomain_GangOfSix_ADRC2_b0}).

\item
Decrease $b_0$ to bring the closed-loop behavior closer to the intended design value, but only as long as the resulting noise sensitivity of the the control signal is tolerable and oscillations do not appear (cf.\ purple plots in \refFig{fig:FrequencyDomain_GangOfSix_ADRC2_b0}).
\end{itemize}

Further properties of $b_0$ and its role were extensively studied in \cite{Zhou:2019}, also for the case of a generalized ADRC incorporating the full plant model instead of the integrator chain approach of conventional linear ADRC.


\subsection{Influence of an additional plant zero}
\label{sec:FrequencyDomain_Influence_Zero}

The gang-of-six approach can also be used to analyze the behavior in cases the actual plant exhibits dynamics beyond the simplified modeling approach of ADRC. As an example with practical relevance, the case of an additional plant zero (right half plane (RHP) or left half plane) will be examined here. This was not covered yet by previous studies such as \cite{Herbst:2013} (which, however, already reported on ADRC's performance when faced with varying plant parameters such as damping and gain). RHP zeros occur, for instance, in boost or buck/boost-derived DC-DC converters in the power electronics application domain \cite{Suntio:2017}.

As visible in \refFig{fig:FrequencyDomain_GangOfSix_ADRC2_Zero}, especially RHP zeros will induce medium-frequency oscillations, which will be visible in the control signal earlier and more pronounced than in the controlled variable. When becoming too dominant, RHP zeros will finally render the closed loop unstable. On the other hand, unmodeled LHP zeros are quite well tolerated in the range examined here.


\section{Discrete-time transfer function implementation}
\label{sec:DiscreteTransferFunction}


\subsection{Discretization of the extended state observer}
\label{sec:DiscreteTransferFunction_ESO}

To establish the notation and introduce the design equations, a brief summary of discrete-time ADRC will be given here. To obtain discrete-time ADRC, only the observer equations must be discretized from their continuous-time counterparts \cite{Herbst:2013}. The state-of-the-art approach for discretizing the extended state observer (ESO) is to employ a current observer approach \cite{Miklosovic:2006}:

\begin{equation}
\Vector{\hat{x}}(k) = \Matrix{A}_\mathrm{ESO} \cdot \Vector{\hat{x}}(k-1) + \Vector{b}_\mathrm{ESO} \cdot u(k-1) + \Vector{l} \cdot y(k)
\label{eqn:ADRC_Discrete_Observer}
\end{equation}
with
\begin{equation}
\begin{aligned}
\Matrix{A}_\mathrm{ESO} &= \Matrix{A}_\mathrm{d} - \Vector{l} \cdot \Vector{c}^\Transpose_\mathrm{d} \cdot \Matrix{A}_\mathrm{d}  \\
\Vector{b}_\mathrm{ESO} &= \Vector{b}_\mathrm{d} - \Vector{l} \cdot \Vector{c}^\Transpose_\mathrm{d} \cdot \Vector{b}_\mathrm{d}
\end{aligned}
\label{eqn:ADRC_Discrete_Observer_AB}
\end{equation}
where $\Matrix{A}_\mathrm{d}$, $\Vector{b}_\mathrm{d}$, $\Vector{c}^\Transpose_\mathrm{d}$ are obtained via zero-order-hold (ZOH) discretization with a sample time $T$ of $\Matrix{A}$, $\Vector{b}$, $\Vector{c}^\Transpose$ from \refEq{eqn:ADRC_A_B_C}:
\begin{equation*}
\Matrix{A}_\mathrm{d} = \Matrix{I} + \sum_{i = 1}^\infty \D\frac{\Matrix{A}^i \cdot T^i}{i!}
,\quad
\Vector{b}_\mathrm{d} = \left( \sum_{i = 1}^\infty \D\frac{\Matrix{A}^{i-1} \cdot T^i}{i!} \right) \cdot \Vector{b}
,\quad
\Vector{c}^\Transpose_\mathrm{d} = \Vector{c}^\Transpose
\end{equation*}

While the controller gains in $\Vector{k}^\Transpose$ do not change, the observer design must be adapted to the discrete-time case. Following the bandwidth parameterization approach, we want to place all observer poles (i.\,e.\ the eigenvalues of $\Matrix{A}_\mathrm{ESO}$) at $z_\mathrm{ESO} = \mathrm{e}^{- k_\mathrm{ESO} \omega_\mathrm{CL} T}$ in the $z$-plane. Therefore the design approach for obtaining the observer gains $\Vector{l}$ is given as:
\begin{equation}
\det\left( z \Matrix{I} - \Matrix{A}_\mathrm{ESO} \right)
\stackrel{!}{=} \left( z - z_\mathrm{ESO} \right)^{n+1}
\label{eqn:ADRC_Design_L_Approach_Discrete}
\end{equation}

\begin{figure}
    \centering%
    \includegraphics[width=\linewidth]{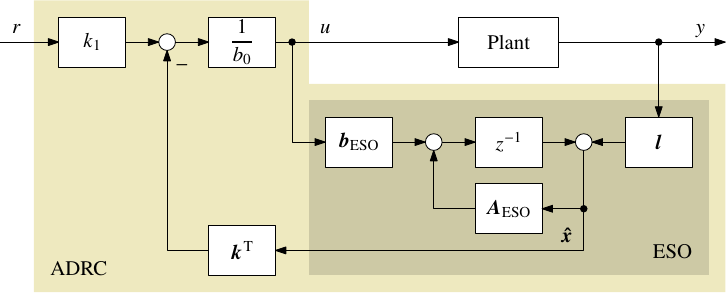}%
    \caption{Discrete-time state-space implementation of linear ADRC using \refEq{eqn:ADRC_Discrete_Observer} and \refEq{eqn:ADRC_Discrete_Observer_AB} for the extended state observer (ESO).}
    \label{fig:ADRC_StateSpace_Discrete}
\end{figure}

For the cases most relevant to industrial practice ($n = 1$ and $n = 2$), the resulting observer gains are given in \refTable{table:ControllerObserverGains_Discrete}. A visualization of the discretized control law is presented in \refFig{fig:ADRC_StateSpace_Discrete}, which is the discrete-time counterpart of the continuous-time case in \refFig{fig:ADRC_StateSpace}.

\subsection{Derivation of discrete-time transfer functions}
\label{sec:DiscreteTransferFunction_Generic}

We follow a similar approach as in \refSec{sec:TransferFunction_Approach} to obtain the transfer function representation of discrete-time ADRC based on the current observer approach from \refSec{sec:DiscreteTransferFunction_ESO}, starting with the $z$-transform of control law and observer dynamics:
\begin{gather}
u(z)
= \frac{1}{b_0} \cdot \left( k_1 r(z) - \Vector{k}^\Transpose \Vector{\hat{x}}(z) \right)
\label{eqn:ADRC_ControlLaw_z}
\\
\Vector{\hat{x}}(z) = z^{-1} \Matrix{A}_\mathrm{ESO} \Vector{\hat{x}}(z) + z^{-1} \Vector{b}_\mathrm{ESO} u(z) + \Vector{l} y(z)
\label{eqn:ADRC_Discrete_Observer_z}
\end{gather}

Putting \refEq{eqn:ADRC_ControlLaw_z} in \refEq{eqn:ADRC_Discrete_Observer_z} yields the closed-loop dynamics of the discrete-time observer:
\begin{gather}
\Vector{\hat{x}}(z)
=
\Matrix{\Phi}_\mathrm{ESO} \cdot
\left( z^{-1} \frac{k_1}{b_0} \Vector{b}_\mathrm{ESO} r(z) + \Vector{l} y(z) \right)
\label{eqn:ADRC_Discrete_Observer_z_ClosedLoop}
\\
\text{with}\quad
\Matrix{\Phi}_\mathrm{ESO}
=
\left(
    \Matrix{I} - z^{-1} \cdot \left( \Matrix{A}_\mathrm{ESO} - \frac{1}{b_0} \Vector{b}_\mathrm{ESO} \Vector{k}^\Transpose \right)
\right)^{-1}
\label{eqn:ADRC_Discrete_Observer_Abbrev}
\end{gather}

Now we can put \refEq{eqn:ADRC_Discrete_Observer_z_ClosedLoop} back in \refEq{eqn:ADRC_ControlLaw_z} to obtain the control law including the closed-loop observer dynamics:
\begin{equation}
u(z)
=
\frac{k_1}{b_0} r(z)
- \frac{1}{b_0} \Vector{k}^\Transpose \cdot
\Matrix{\Phi}_\mathrm{ESO}
\cdot
\left( z^{-1} \frac{k_1}{b_0} \Vector{b}_\mathrm{ESO} r(z) + \Vector{l} y(z) \right)
\label{eqn:ControlLaw_Discrete}
\end{equation}

\begin{figure}
    \centering%
    \includegraphics[width=\linewidth]{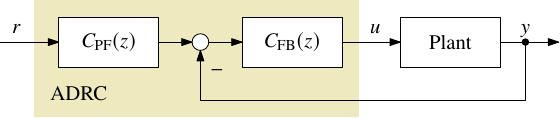}%
    \caption{Control loop with discrete-time transfer function based ADRC implementation consisting of feedback controller $C_\mathrm{FB}(z)$ and reference signal prefilter $C_\mathrm{PF}(z)$.}
    \label{fig:ADRC_TransferFunction_Discrete}
\end{figure}

In constrast to continuous-time case, which required three transfer functions for a realizable representation of the control law (as introduced in \refSec{sec:TransferFunction_Approach}), discrete-time discrete-time ADRC can be implemented using using only two transfer functions: a feedback controller $C_\mathrm{FB}(z)$, and a reference signal prefilter $C_\mathrm{PF}(z)$ (cf.\ \refFig{fig:ADRC_TransferFunction_Discrete}). The approach for the control law is:
\begin{equation}
u(z) = C_\mathrm{FB}(z) \cdot \left[ C_\mathrm{PF}(z) \cdot r(z) - y(z) \right]
\label{eqn:Transfer_Uz_Generic}
\end{equation}

By comparing \refEq{eqn:ControlLaw_Discrete} and \refEq{eqn:Transfer_Uz_Generic}, we can obtain $C_\mathrm{FB}(z)$ and $C_\mathrm{PF}(z)$:
\begin{gather*}
C_\mathrm{FB}(z)
= \frac{1}{b_0} \Vector{k}^\Transpose \cdot \Matrix{\Phi}_\mathrm{ESO} \cdot \Vector{l}
\\
C_\mathrm{PF}(z)
= \frac{ \frac{k_1}{b_0} \cdot \left( 1 - z^{-1} \Vector{k}^\Transpose \cdot \Matrix{\Phi}_\mathrm{ESO} \cdot \frac{1}{b_0} \Vector{b}_\mathrm{ESO} \right) }{ C_\mathrm{FB}(z) }
\end{gather*}

For a discrete-time implementation of $n$-th order ADRC, these two transfer functions can be represented using the following coefficients:

\begin{equation*}
C_{\mathrm{FB},n}(z) =
\frac
{ \D\sum_{i = 0}^n \beta_i z^{-i} }
{ 1 + \D\sum_{i = 1}^{n+1} \tilde{\alpha}_i z^{-i} }
,\quad
C_{\mathrm{PF},n}(z) =
\frac
{ \D\sum_{i = 0}^{n+1} \gamma_i z^{-i} }
{ 1 + \D\frac{1}{\beta_0} \sum_{i = 1}^n \beta_i z^{-i} }
.
\end{equation*}

Note that the $z^{-1}$-based numerator polynomial of $C_{\mathrm{PF},n}(z)$ is of order $(n+1)$ compared to order $n$ in the denominator. What was a realizability hurdle in the continuous-time case poses no problem in a digital filter implementation, hence only two instead of three transfer functions are necessary here.

Moreover, note that the denominator coefficients of the feedback controller $C_{\mathrm{FB},n}(z)$ are given as $\tilde{\alpha}_i$. This shall indicate an intermediate result, since a modification of the feedback controller with additional benefits will be introduced in the following section.


\subsection{Feedback controller with factored-out accumulator}
\label{sec:DiscreteTransferFunction_ExtAcc}

In a transfer function implementation of an $n$-th order ADRC, the feedback controller transfer function includes an integrator. Therefore $C_{\mathrm{FB},n}(z)$ will have a pole at $z = 1$, and we can factor this discrete-time accumulator out:
\begin{align}
C_{\mathrm{FB},n}(z)
&=
\frac
{ \D\sum_{i = 0}^n \beta_i z^{-i} }
{ 1 + \D\sum_{i = 1}^{n+1} \tilde{\alpha}_i z^{-i} }
=
\frac
{ \D\sum_{i = 0}^n \beta_i z^{-i} }
{ 1 + \D\sum_{i = 1}^n \alpha_i z^{-i} }
\cdot \frac{1}{1 - z^{-1}}
\notag\\
&=
\Delta C_{\mathrm{FB},n}(z) \cdot \frac{1}{1 - z^{-1}}
\label{eqn:Transfer_Cz_FB_Acc}
\end{align}

For the $\alpha$ and $\tilde{\alpha}$ parameters in \refEq{eqn:Transfer_Cz_FB_Acc}, the following relation can be given:
\begin{equation}
\alpha_i = 1 + \D\sum_{j = 1}^{i} \tilde{\alpha}_j
\label{eqn:Transfer_Alpha}
\end{equation}

\begin{proof}
If we put $z = 1$ in the denominator of \refEq{eqn:Transfer_Cz_FB_Acc}, we obtain $1 + \sum_{i = 1}^{n+1} \tilde{\alpha}_i = 0$, since $C_{\mathrm{FB},n}(z)$ contains an integrator (pole at $z = 1$). Therefore $\tilde{\alpha}_{n+1} = - \left(1 + \sum_{i = 1}^{n} \tilde{\alpha}_i\right)$ holds, which allows us to prove \refEq{eqn:Transfer_Alpha} by putting it back into the denominator of \refEq{eqn:Transfer_Cz_FB_Acc} and proceed as follows:
\begingroup
\allowdisplaybreaks
\begin{align*}
&\left( 1 + \D\sum_{i = 1}^n \alpha_i z^{-i} \right)
\cdot \left( 1 - z^{-1} \right)
\\
&=
\left( 1 + \D\sum_{i = 1}^n \left( 1 + \D\sum_{j = 1}^{i} \tilde{\alpha}_j \right) \cdot z^{-i} \right)
\cdot \left( 1 - z^{-1} \right)
\\
&=
1 + \left( 1 + \tilde{\alpha}_1 \right) \cdot z^{-1}
+ \D\sum_{i = 2}^n \left( 1 + \D\sum_{j = 1}^{i} \tilde{\alpha}_j \right) \cdot z^{-i}
- z^{-1}
\\
&\phantom{=} - \D\sum_{i = 2}^{n+1} \left( 1 + \D\sum_{j = 1}^{i-1} \tilde{\alpha}_j \right) \cdot z^{-i}
\\
&=
1 + \D\sum_{i = 1}^n \tilde{\alpha}_i z^{-i}
- \Bigg( \underbrace{  1 + \D\sum_{j = 1}^{n} \tilde{\alpha}_j }_{-\tilde{\alpha}_{n+1}} \Bigg) \cdot z^{-(n+1)}
=
1 + \D\sum_{i = 1}^{n+1} \tilde{\alpha}_i z^{-i}
.
\label{eqn:Transfer_Alpha_Proof}
\end{align*}
\endgroup
\end{proof}

The benefits of implementing the feedback controller transfer function $C_{\mathrm{FB},n}(z)$ with a factored-out accumulator are:
\begin{itemize}
\item
Compared to a state-space implementation of ADRC, fewer multiplications in the control law are required, decreasing the computational burden of the control algorithm to be implemented (first-order: 7 instead of 11, second-order: 11 instead of 19; and this is already using the matrices given in \refEq{eqn:ADRC_Discrete_Observer_AB} for the observer \refEq{eqn:ADRC_Discrete_Observer}). Additionally, factoring out the accumulator saves one $\alpha$ coefficient to be stored for the transfer function implementation.

\item
With the factored-out accumulator, a straightforward solution is possible for clamping the controller output: the accumulator has to be simply replaced by a clamped accumulator, thereby avoiding windup issues \cite{Astrom:2006}.

\item
The use of one common accumulator allows to switch between different controllers easily \cite{Astrom:2006}, for example when using an ``override control'' scheme \cite{Smith:2010} with two controllers acting on the same plant input.

\item
A better equivalence to the continuous-time case is achieved, where the feedback controller transfer function includes a factored-out integrator as well, cf.\ \refEq{eqn:Transfer_C_FB_PF_FF_n}.
\end{itemize}


\subsection{Summary}
\label{sec:DiscreteTransferFunction_Summary}

Discrete-time linear ADRC can be implemented using transfer functions with the following control law,
\begin{equation*}
u(z) = C_\mathrm{FB}(z) \cdot \left[ C_\mathrm{PF}(z) \cdot r(z) - y(z) \right]
,
\end{equation*}
where the feedback controller transfer function $C_{\mathrm{FB}}(z)$ and the reference signal prefilter $C_{\mathrm{PF}}(z)$ for $n$-th order ADRC are given as:

\begin{align*}
C_{\mathrm{FB}}(z) &=
\frac
{ \D\sum_{i = 0}^n \beta_i z^{-i} }
{ 1 + \D\sum_{i = 1}^n \alpha_i z^{-i} }
\cdot \frac{1}{1 - z^{-1}}
,
\\
C_{\mathrm{PF}}(z) &=
\frac
{ \D\sum_{i = 0}^{n+1} \gamma_i z^{-i} }
{ 1 + \D\frac{1}{\beta_0} \sum_{i = 1}^n \beta_i z^{-i} }
.
\end{align*}

For first- and second-order ADRC, the discrete-time transfer functions are given again in detail in \refTable{table:Transfer_Cz}, while the according $\alpha$, $\beta$, and $\gamma$ coefficients can be found in \refTable{table:TF_Parameters1_Discrete} and \refTable{table:TF_Parameters2_Discrete}.

\begin{table*}
\begin{center}
\caption{Discrete-time transfer function implementation of first- and second-order ADRC. The $\alpha$, $\beta$, $\gamma$ coefficients for the first- and second order case can be obtained from \refTable{table:TF_Parameters1_Discrete} and \refTable{table:TF_Parameters2_Discrete}, respectively.}
\label{table:Transfer_Cz}
\begin{tabular*}{\linewidth}{@{\extracolsep\fill}rll@{\extracolsep\fill}}
    \toprule
    \textbf{}  &  \textbf{Feedback controller}  &  \textbf{Reference signal prefilter}  \\
    \midrule
    \\[-0.9em]
    \textbf{First-order ADRC}
    &  $\D C_{\mathrm{FB},1}(z) = \frac{ \beta_0 + \beta_1 z^{-1} }{ 1 + \alpha_1 z^{-1} } \cdot \frac{1}{1 - z^{-1}}$
    &  $\D C_{\mathrm{PF},1}(z) = \frac{ \gamma_0 + \gamma_1 z^{-1} + \gamma_2 z^{-2} }{ 1 + \frac{\beta_1}{\beta_0} z^{-1} }$
    \\[2.0em]
    \textbf{Second-order ADRC}
    &  $\D C_{\mathrm{FB},2}(z) = \frac{ \beta_0 + \beta_1 z^{-1} + \beta_2 z^{-2} }{ 1 + \alpha_1 z^{-1} + \alpha_2 z^{-2} } \cdot \frac{1}{1 - z^{-1}}$
    &  $\D C_{\mathrm{PF},2}(z) = \frac{ \gamma_0 + \gamma_1 z^{-1} + \gamma_2 z^{-2} + \gamma_3 z^{-3} }{ 1 + \frac{\beta_1}{\beta_0} z^{-1} + \frac{\beta_2}{\beta_0} z^{-2} }$
    \\[1.5em]
    \bottomrule
\end{tabular*}
\end{center}
\end{table*}

\begin{table*}
\begin{center}
\caption{Discrete-time transfer function parameters for first-order ADRC, given in terms of the controller and discrete-time observer gains (see \refTable{table:ControllerObserverGains} and \refTable{table:ControllerObserverGains_Discrete}, respectively), and in terms of bandwidth tuning parameters. The latter are $\omega_\mathrm{CL}$ (desired closed-loop bandwidth) and $z_\mathrm{ESO} = \mathrm{e}^{- k_\mathrm{ESO} \omega_\mathrm{CL} T}$ (observer pole locations in $z$-domain, based on the observer bandwidth factor $k_\mathrm{ESO}$). Common parameters are $b_0$ (gain parameter of the plant model) and $T$ (sample time of the discretized implementation).}
\label{table:TF_Parameters1_Discrete}
\begin{tabular*}{\linewidth}{@{\extracolsep\fill}rll@{\extracolsep\fill}}
    \toprule
    \textbf{Parameter}  &  \textbf{General terms}  &  \textbf{Bandwidth parameterization}  \\
    \midrule
    \\[-0.9em]
    $\alpha_1$
    &   $(T k_1 - 1) \cdot (1 - l_1)$
    &   $\D\left( T \omega_\mathrm{CL} - 1 \right) \cdot z_\mathrm{ESO}^2$
    \\[1.0em]
    $\beta_0$
    &   $\D\frac{1}{b_0} \cdot \left( k_1 l_1 + l_2 \right)$
    &   $\D\frac{ 1 }{ b_0 T } \cdot \left[ T \omega_\mathrm{CL} \cdot \left(1 - z_\mathrm{ESO}^2 \right) + \left(1 - z_\mathrm{ESO}\right)^2 \right]$
    \\[1.0em]
    $\beta_1$
    &   $\D\frac{1}{b_0} \cdot \left( T k_1 l_2 - k_1 l_1 - l_2 \right)$
    &   $\D\frac{ 1 }{ b_0 T } \cdot \left[ -2 T \omega_\mathrm{CL} z_\mathrm{ESO} \cdot \left(1 - z_\mathrm{ESO} \right) - \left(1 - z_\mathrm{ESO}\right)^2 \right]$
    \\[1.2em]
    $\gamma_0$
    &   $\D\frac{k_1}{k_1 l_1 + l_2}$
    &   $\D\frac{ T \omega_\mathrm{CL} }{ T \omega_\mathrm{CL} \cdot \left(1 - z_\mathrm{ESO}^2 \right) + \left(1 - z_\mathrm{ESO}\right)^2 }$
    \\[1.5em]
    $\gamma_1$
    &   $\D\frac{k_1 \cdot (T l_2 + l_1 - 2)}{k_1 l_1 + l_2}$
    &   $\D\frac{ -2 T \omega_\mathrm{CL} z_\mathrm{ESO} }{ T \omega_\mathrm{CL} \cdot \left(1 - z_\mathrm{ESO}^2 \right) + \left(1 - z_\mathrm{ESO}\right)^2 }$
    \\[1.5em]
    $\gamma_2$
    &   $\D\frac{k_1 \cdot (1 - l_1)}{k_1 l_1 + l_2}$
    &   $\D\frac{ T \omega_\mathrm{CL} z_\mathrm{ESO}^2 }{ T \omega_\mathrm{CL} \cdot \left(1 - z_\mathrm{ESO}^2 \right) + \left(1 - z_\mathrm{ESO}\right)^2 }$
    \\[1.5em]
    \bottomrule
\end{tabular*}
\end{center}
\end{table*}


\setlength\rotFPtop{250pt}
\begin{sidewaystable*}
\begin{center}
\caption{Discrete-time transfer function parameters for second-order ADRC, given in terms of the controller and discrete-time observer gains (see \refTable{table:ControllerObserverGains} and \refTable{table:ControllerObserverGains_Discrete}, respectively), and in terms of bandwidth tuning parameters. The latter are $\omega_\mathrm{CL}$ (desired closed-loop bandwidth) and $z_\mathrm{ESO} = \mathrm{e}^{- k_\mathrm{ESO} \omega_\mathrm{CL} T}$ (observer pole locations in $z$-domain, based on the observer bandwidth factor $k_\mathrm{ESO}$). Common parameters are $b_0$ (gain parameter of the plant model) and $T$ (sample time of the discretized implementation).}
\label{table:TF_Parameters2_Discrete}
\begin{tabular*}{\textwidth}{@{\extracolsep\fill}rll@{\extracolsep\fill}}
    \toprule
    \textbf{Parameter}  &  \textbf{General terms}  &  \textbf{Bandwidth parameterization}  \\
    \midrule
    \\[-0.9em]
    $\alpha_1$
    &   $\D\frac{T^2}{2} \cdot \left(k_1 - k_1 l_1 - k_2 l_2\right) + T k_2 + T l_2 + l_1 - 2$
    &   $\D \frac{1}{2} \cdot \Big[ T^2 \omega_\mathrm{CL}^2 z_\mathrm{ESO}^3 + T \omega_\mathrm{CL} \cdot \left( 1 + 3 z_\mathrm{ESO} + 3 z_\mathrm{ESO}^2 - 3 z_\mathrm{ESO}^3 \right) + \left( 1 - 3 z_\mathrm{ESO} - 3 z_\mathrm{ESO}^2 + z_\mathrm{ESO}^3 \right) \Big]$
    \\[1.0em]
    $\alpha_2$
    &   $\D \left( \frac{T^2 k_1}{2} - T k_2 + 1 \right) \cdot \left( 1 - l_1 \right)$
    &   $\D \frac{z_\mathrm{ESO}^3}{2} \cdot \left(T^2 \omega_\mathrm{CL}^2 - 4 T \omega_\mathrm{CL} + 2 \right)$
    \\[1.5em]
    $\beta_0$
    &   $\D\frac{1}{b_0} \cdot \left[ k_1 l_1 + k_2 l_2 + l_3 \right]$
    &   $\D\frac{ 1 }{ b_0 T^2 } \cdot \left[ T^2 \omega_\mathrm{CL}^2 \cdot \left( 1 - z_\mathrm{ESO}^3 \right) + 3 T \omega_\mathrm{CL} \cdot \left( 1 - z_\mathrm{ESO} - z_\mathrm{ESO}^2 + z_\mathrm{ESO}^3 \right) + \left( 1 - z_\mathrm{ESO} \right)^3 \right]$
    \\[1.0em]
    $\beta_1$
    &   $\D\frac{1}{b_0} \cdot \left[ \frac{T^2 k_1 l_3}{2} + T k_1 l_2 + T k_2 l_3 - 2 ( k_1 l_1 + k_2 l_2 + l_3 ) \right]$
    &   $\D\frac{ 1 }{ b_0 T^2 } \cdot \left[ -3 T^2 \omega_\mathrm{CL}^2 z_\mathrm{ESO} \cdot \left( 1 - z_\mathrm{ESO}^2 \right) - 4 T \omega_\mathrm{CL} \cdot \left( 1 - 3 z_\mathrm{ESO}^2 + 2 z_\mathrm{ESO}^3 \right) - 2 \cdot \left( 1 - z_\mathrm{ESO} \right)^3 \right]$
    \\[1.0em]
    $\beta_2$
    &   $\D\frac{1}{b_0} \cdot \left[ \frac{T^2 k_1 l_3}{2} - T k_1 l_2 - T k_2 l_3 + \left( k_1 l_1 + k_2 l_2 + l_3 \right) \right]$
    &   $\D\frac{ 1 }{ b_0 T^2 } \cdot \left[ 3 T^2 \omega_\mathrm{CL}^2 z_\mathrm{ESO}^2 \cdot \left( 1 - z_\mathrm{ESO} \right) + T \omega_\mathrm{CL} \cdot \left( 1 + 3 z_\mathrm{ESO} - 9 z_\mathrm{ESO}^2 + 5 z_\mathrm{ESO}^3 \right) + \left( 1 - z_\mathrm{ESO} \right)^3 \right]$
    \\[1.5em]
    $\gamma_0$
    &   $\D\frac{k_1}{k_1 l_1 + k_2 l_2 + l_3}$
    &   $\D\frac{ T^2 \omega_\mathrm{CL}^2 }{ T^2 \omega_\mathrm{CL}^2 \cdot \left( 1 - z_\mathrm{ESO}^3 \right) + 3 T \omega_\mathrm{CL} \cdot \left( 1 - z_\mathrm{ESO} - z_\mathrm{ESO}^2 + z_\mathrm{ESO}^3 \right) + \left( 1 - z_\mathrm{ESO} \right)^3 }$
    \\[1.5em]
    $\gamma_1$
    &   $\D\frac{k_1 \cdot (T^2 l_3 + 2 T l_2 + 2 l_1 - 6)}{k_1 l_1 + k_2 l_2 + l_3}$
    &   $\D\frac{ -3 T^2 \omega_\mathrm{CL}^2 z_\mathrm{ESO} }{ T^2 \omega_\mathrm{CL}^2 \cdot \left( 1 - z_\mathrm{ESO}^3 \right) + 3 T \omega_\mathrm{CL} \cdot \left( 1 - z_\mathrm{ESO} - z_\mathrm{ESO}^2 + z_\mathrm{ESO}^3 \right) + \left( 1 - z_\mathrm{ESO} \right)^3 }$
    \\[1.5em]
    $\gamma_2$
    &   $\D\frac{k_1 \cdot (T^2 l_3 - 2 T l_2 - 4 l_1 + 6)}{k_1 l_1 + k_2 l_2 + l_3}$
    &   $\D\frac{ 3 T^2 \omega_\mathrm{CL}^2 z_\mathrm{ESO}^2 }{ T^2 \omega_\mathrm{CL}^2 \cdot \left( 1 - z_\mathrm{ESO}^3 \right) + 3 T \omega_\mathrm{CL} \cdot \left( 1 - z_\mathrm{ESO} - z_\mathrm{ESO}^2 + z_\mathrm{ESO}^3 \right) + \left( 1 - z_\mathrm{ESO} \right)^3 }$
    \\[1.5em]
    $\gamma_3$
    &   $\D\frac{k_1 \cdot (l_1 - 1)}{k_1 l_1 + k_2 l_2 + l_3}$
    &   $\D\frac{ - T^2 \omega_\mathrm{CL}^2 z_\mathrm{ESO}^3 }{ T^2 \omega_\mathrm{CL}^2 \cdot \left( 1 - z_\mathrm{ESO}^3 \right) + 3 T \omega_\mathrm{CL} \cdot \left( 1 - z_\mathrm{ESO} - z_\mathrm{ESO}^2 + z_\mathrm{ESO}^3 \right) + \left( 1 - z_\mathrm{ESO} \right)^3 }$
    \\[1.5em]
    \bottomrule
\end{tabular*}
\end{center}
\end{sidewaystable*}

\section{Conclusion}

To foster a more widespread adoption of linear ADRC in industrial practice, this article has made three contributions.

Firstly, a representation of ADRC using realizable transfer functions was introduced. This allows to compare ADRC to an existing ``classical'' solution more easily, for example by analyzing the pole/zero placement of the feedback controller part.

Secondly, based on the transfer function implementation, a detailed frequency-domain analysis of linear ADRC was performed, including plant modeling and the feedback controller part. A frequency-domain inspection of the tuning parameter impact on the ``gang-of-six'' transfer functions of a closed loop with ADRC supports the tuning process, e.\,g.\ when looking for a compromise between low-frequency disturbance rejection and high-frequency noise sensitivity.

Finally, an exact and low-footprint transfer function realization of discrete-time linear ADRC was derived, with ready-to-use coefficient tables given for first- and second-order ADRC---the most relevant contenders to existing PI and PID controller solutions. With the results presented in this article, an efficient implementation of linear ADRC even on low-cost target hardware should have become a low-effort and straightforward task.


{
\newpage
\color{black!50}%
\printendnotes
}

\end{document}